# A node-charge graph-based online carshare rebalancing policy with capacitated electric charging

Theodoros P. Pantelidis[1], Li Li[1,2], Tai-Yu Ma[3], Joseph Y. J. Chow[1*], Saif Eddin G. Jabari[2]

[1] Department of Civil & Urban Engineering, New York University, Brooklyn, NY, USA
[2] Division of Engineering, New York University Abu Dhabi, Abu Dhabi, UAE
[3] Luxembourg Institute of Socio-Economic Research, Luxembourg
[*]Corresponding Author Email: joseph.chow@nyu.edu

## Abstract

Viability of electric car-sharing operations depends on rebalancing algorithms. Earlier methods in the literature suggest a trend toward non-myopic algorithms using queueing principles. We propose a new rebalancing policy using cost function approximation. The cost function is modeled as a p-median relocation problem with minimum cost flow conservation and path-based charging station capacities on a static node-charge graph structure. The cost function is NP-complete, so a heuristic is proposed that ensures feasible solutions that can be solved in an online system. The algorithm is validated in a case study of electric carshare in Brooklyn, New York, with demand data shared from BMW ReachNow operations in September 2017 (262 vehicle fleet, 231 pickups per day, 303 traffic analysis zones (TAZs)) and charging station location data (18 charging stations with 4 port capacities). The proposed non-myopic rebalancing heuristic reduces the cost increase compared to myopic rebalancing by 38%. Other managerial insights are further discussed.

# 1. Introduction

Car sharing operations form an essential part of "smart mobility" solutions in congested cities. According to Martin and Shaheen (2016), a single carshare vehicle can replace 7 to 11 personal vehicles on the road, or between 5 to 20 vehicles by other accounts (Navigant Research, 2017). The common practice in such services is to book specific time slots and reserve a vehicle from a specific location. The return location is required to be the same for "two-way" systems but is relaxed for "one-way" systems. Examples of free-floating systems are the BMW ReachNow car sharing system in Brooklyn (until 2018) and Car2Go in New York City, with service areas in 2017 shown in Figure 1.

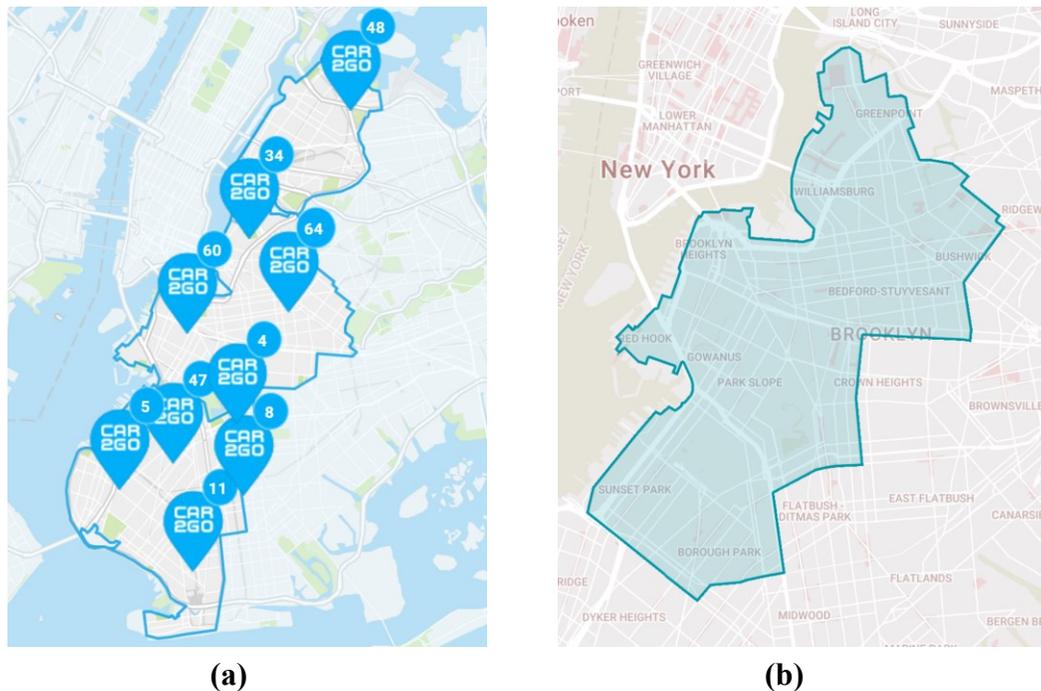

(a)          (b)
**Figure 1.** Examples of free-floating carshare systems: (a) Car2Go (source: car2go.com) and (b) BMW ReachNow (source: reachnow.com).

In large car sharing systems, vehicle rebalancing is one of the primary challenges to ensuring efficiency and providing an adequate level of service. Potential customers may end up waiting or accessing a farther location, or even balk from using the service, if there is no available vehicle within a reasonable proximity (which may involve substantial access, e.g. taking a subway from downtown Manhattan to midtown to pick up a car) or no parking or return location available near the destination. Rebalancing involves having either the system staff or users (through incentives) periodically drop off vehicles at locations that would better match supply to demand (see Nourinejad et al., 2015). Inefficient operations may cause systems to be shut down (Krok, 2016).

Car-sharing companies have further considered electric vehicle (EV) fleets to be more sustainable to reduce gasoline consumption. Not only do Zipcar, Car2Go, and ReachNow all operate some EVs in their worldwide fleets (car2go in 2019: Stuttgart, Madrid and Amsterdam), some startup carsharing businesses rely exclusively on EVs: e.g. Autolib' and Cité Lib in France; BlueIndy, DriveNow in Copenhagen; Carma in San Francisco; and Los Angeles' Low-Income

Plan for EV car sharing (Lufkin, 2016). With a total fleet of approximately 1,400 EVs, car2go is one of the largest providers in the electric vehicle carsharing sector.

Vehicle rebalancing efficiency is further hampered in an EV environment. EV fleets face the added challenge of limited availability of fast charging infrastructure (as of 2019 there are seven fast DC public charging stations in Manhattan including Tesla stations (Chargehub, 2019)), which still take longer to recharge (~ 30 minutes) than gasoline vehicles. While there is an abundant literature on methods to handle carshare rebalancing, research on rebalancing EVs under capacitated charging stations is limited. There is no model formulated yet for one-way EV carsharing rebalancing that captures all the following: 1) the stochastic dynamic nature of rebalancing with stochastic demand; 2) incorporating users' access cost to vehicles; and 3) capacities at EV charging stations.

The problem is inherently a type of Markov decision process (see Powell, 2011; Chow and Sayarshad, 2016; Sayarshad and Chow, 2017) requiring an optimal rebalancing policy. We propose a policy based on cost function approximation (CFA) which uses a novel graph structure that allows the three challenges to be addressed. The resulting cost function is a p-median relocation model that extends the rebalancing model of Sayarshad and Chow (2017). We call this method the "node-charge graph" approach as it involves expanding the network to a discretized charge level dimension. The approach is similar to Zhang et al. (2019) who employ a "space-time-battery" graph approach. However, their method faces computational challenges with the additional time dimension, especially if the carsharing system involves charging activities in the order of minutes while relocation intervals and vehicle reservations are made on the order of hours. To be fair though, Zhang et al. (2019) study an offline problem focusing on evaluating system equilibration, which differs from the operational objective of this study. We present a more elegant modeling framework that eschews the time dimension to allow for online application.

By using the model from Sayarshad and Chow (2017) as a basis, the online relocation cost function can be calibrated to account for the stochastic demand using queue delay approximation. Unlike other queue delay-based approaches that assume demand at one node is only assigned to the servers at that same node, our delays are directly incorporated into a facility relocation problem that more realistically allows for, and minimizes, access from one node to another. The resulting model is a cost function for a relocation policy with minimum cost flow constraints as opposed to the conventional transportation problem constraints. A novel heuristic is proposed to determine the relocation policy in an online setting. We implement the algorithm in a proprietary agent-based simulation (see Li et al., 2019) to test its performance over a time horizon against benchmark algorithms using demand data from BMW ReachNow carshare operations in Brooklyn in 2017.

The paper is organized as follows. Section 2 presents the literature review. Section 3 presents the proposed graph structure, model formulation, and proposed heuristic. Section 4 presents computational tests to verify the model and heuristic effectiveness. Section 5 presents the agent simulation experiment design and case study of the Brooklyn carshare fleet to validate the heuristic performance. Section 6 concludes this study.

## 2. Literature review

### 2.1. Rebalancing literature

Optimal rebalancing considers trade-offs between vehicle availability for a given random demand distribution, access distance, and relocation costs. For EV systems, there are further trade-offs involving charging duration, charging station capacity, and demand distribution over different charge levels.

Early studies for designing car-sharing systems relied on simulation for evaluation (Barth and Todd, 1999). More systematic mathematical models to optimize car-sharing fleets have since been proposed with rebalancing and related operational challenges, including Fan et al. (2008), Kek et al. (2009), Nair and Miller-Hooks (2011), Di Febbraro et al. (2012), and Sayarshad and Chow (2017). There are also studies on station location with rebalancing (e.g. Chow and Sayarshad, 2014), pricing incentives (Clemente et al., 2014; Jorge et al., 2015; Waserhole and Jost, 2016), parking reservations (Kaspi et al., 2016), routing personnel (Brugl­ieri et al., 2014; Nourinejad et al., 2015), fleet sizing (Hu and Liu, 2016), and integrated multimodal systems (Ma et al., 2019a), among others.

Less focus has been given to methods for rebalancing EV carshare systems, however. A smaller subset of studies emerged in recent years to tackle this heightened challenge. Two general methods have been adopted. The first assumes demand is sufficiently deterministic in a multiperiod setting (e.g. Xu et al., 2018; Gambella et al., 2018; Zhang et al., 2019). This can be problematic for most systems where the demand is not made for repeated commute trips and/or the fleet has a sparse spatial distribution. Vehicles are also assumed to be picked up at one location and directly dropped off at a destination, which is not typically the case in carsharing as customers may run multiple errands in a trip chain or leaving the service coverage area entirely before returning to drop off a vehicle.

The second group of methods assumes stochastic demand, either through stochastic programming (Brandstätter et al., 2017), simulation (Boyacı et al., 2015), or with Markovian demand (Li et al., 2016). The latter Markovian queueing models appear promising, but earlier EV studies either assume a simplistic relocation policy (Li et al., 2016) or, in the case of queueing networks (Waserhole and Jost, 2016), require customers to pick up vehicles only at the same zone/node.

Discrete network approaches make use of queueing to handle the uncertainty in stochastic demand. However, these approaches have not been used for EV charging settings either. Existing approaches include (1) queueing network models like Waserhole and Jost (2016) and Zhang and Pavone (2016), which require demand to be served by vehicles only from the same zone; or (2) the queueing-based relocation model from Sayarshad and Chow (2017) and Ma et al. (2019a), which allows vehicles to cover demand at other zones. None of those consider EV demand and charging constraints. It is quite clear that EV car sharing systems present a more complex environment in terms of mathematical modeling and decision analysis that current state-of-the-art methods do not fully address.

## 2.2. Relocation for stochastic demand

Rebalancing policies for carshare should include access costs for customers to allow them to enter a system at one node and pick up a vehicle at a different node. This requires the use of facility location models for the cost functions. For stochastic demand, facility location can use queue delay to anticipate future opportunity costs of a certain location solution. Queue delay is modeled by defining each service node as a queue with $s$ servers, where service entails usage of the vehicle until it is returned to the network (which may be at a different zone). In this case, however, even a simplified assumption of a M/M/s stochastic queue at each zone results in a nonlinear objective. Since nonlinear integer programming problems are undesirable, researchers have proposed alternative methods to handle the queueing.

One such way is the Q-MALP model from Marianov and ReVelle (1996), who showed that the queue delay objective can instead be cast as a set of piecewise linear constraints for the intensity to be within a specified reliability level $\eta$. Because the intensity parameter can be preprocessed for different numbers of servers, it is possible to solve a facility location problem with desired queueing-based service reliability as a mixed integer linear programming problem. The model has since been modified to handle maximal coverage (Marianov and Serra, 1998), server allocation (Marianov and Serra, 2002), and p-median coverage with relocation costs (Sayarshad and Chow, 2017).

Queueing-based facility location models handle everything that the "queueing network" models can (by restricting access thresholds to the same node), and furthermore allow for inter-zonal matching of vehicles to demand. However, the relocation component in Sayarshad and Chow (2017) is based on a bipartite transportation problem of moving excess servers to locations in demand of servers. This is fine for a non-EV carsharing system, but for EV charging the mechanics are more complex because charging trips need to consider battery range, and both proximity and availability of charging stations (see Jung et al., 2014). The model from Sayarshad and Chow (2017) also does not distinguish demand for a minimum charge level less than 100%. For example, a customer should be allowed to request a vehicle with 60% charge or more and be allowed to pick up a vehicle with 80% charge.

We address this issue by first proposing a static node-charge expanded network representation of the location problem for a CFA policy for EV carshare fleets. Under this representation, a network is replicated into multiple charge levels and movement from one charge level up to another represents recharging activity. We further propose the first queueing-based facility relocation problem with minimum cost flow relocation for the cost function, where a vehicle may start at a lower charge level, be repositioned at a charging station, recharge up several charge levels, and then be matched to customers at nearby zones. This is non-trivial because the capacity is not at the link level, i.e. one car charging at a station from 20% to 40% and another car charging from 60% to 80% would occupy different "links" in this graph but be competing for the same capacity. This model is formulated as a mixed integer programming problem. For larger cases or online operation, we propose a novel heuristic algorithm that ensures the three dimensions of feasibility of intermediate solutions: coverage, queueing intensity, and charging station capacity.

## 3. Proposed methodology

### 3.1. Problem statement

The rebalancing problem for a fleet of EVs is a Markov decision process (MDP). A central operator controls periodic rebalancing of a homogeneous fleet of electric vehicles $f \in F$ which are randomly picked up and dropped off across a given network $G(N, A)$ of zone centroids $N$ connected by bi-directional links $A$ over a finite planning horizon $T = \{0, 1, \ldots, |T|\}$. A subset of these zone centroids is designated as charging stations $J \subset N$ with a finite number of chargers $u_j, j \in J$, at which vehicles may recharge or idle when not in service.

The locations and charge levels of the vehicles are known. Customers arrive to the system, book and pick up a desired EV of their choice. When they finish using the vehicle over a duration, the vehicle is dropped back into the network. The pickup and drop-off locations and durations of booking are random with known distributions. Let $S_t = (s_c, s_v)$ denote the system state vector, where $t$ is a discrete time step from $t = 0$ at the beginning of operations. The state vector $s_c(t)$ is a random matrix that contains all corresponding customer attributes in each row for location of pickup and drop-off of a vehicle, pickup and drop-off times, desired battery charge level, and returned vehicle charge level. The pickup location and desired battery charge level are realized at the pickup time while the drop-off location and returned charge level are realized upon drop-off time.

The state vector of vehicle attributes $s_v(t)$ describes the state of each vehicle in the fleet. Each row corresponds to a vehicle $f$ and includes a set of time-dependent attributes that indicate the status of vehicle $f$ (idle, rebalancing, in-service), its current position, and the current battery level. An idle vehicle switches to in-service when it is booked by a customer, and switches from idle to rebalancing when it is assigned to another location by the operator. An in-service vehicle changes to idle when it is returned to the system. A rebalancing vehicle becomes idle when it completes the rebalancing assignment.

The operator only manages recharging and rebalancing of idle vehicles in anticipation of future requests. The objective is to minimize the cumulative operating cost (rebalancing and recharging) and user cost (access time, wait time). If multiple customers are waiting for a vehicle with the required charge level, they are served in first-come-first-served order. The fleet operator only makes two sets of decisions: (1) when and where to rebalance idle vehicles, and (2) their target charge levels. Once a vehicle starts its rebalancing, it will not be cancelled or rescheduled.

Let $F_{idle}$ denote the idle vehicle set at any time step $t$. Let $a_\tau$ denote the action vector of all idle EVs at time step $t$, where $a_{ft}$ is the action of EV $f$. Each idle EV can take one of the following actions:

$$a_{ft} = \begin{cases} Pass: \text{maintains the idle status without charging} \\ Rebalance: \text{rebalance to another node.} \\ Charge: \text{Maintains idle status while recharging} \end{cases}$$

Let $r(S_t, a_t) \in \mathbb{R}$ denote the immediate reward received by taking action vector $a_t$ in state $S_t$. The policy $\pi$ is a sequence of decision rules $(X_0^\pi, X_1^\pi, ..., X_T^\pi)$ where $X_t^\pi$ is a function mapping state $S_t$ in time step $t$ to an action $A_t \in A(S_t)$ comprised of each vehicle's $a_{ft}$. We seek to provide the fleet operator with a policy $\pi^*$ that minimizes the expected total costs incurred throughout the time horizon conditional on the initial state as shown in Eq. (1) (expressed in terms of cost minimization instead of the more conventional payoff maximization for consistency).

$$Z(\pi^*) = \mathbb{E} \min_{\pi \in \Pi} \left[ \sum_{t \in T} C(S_t, X_t^\pi(S_t)) \middle| S_0 \right], \quad (1)$$

where $\Pi$ is the set of all policies. Due to the curse of dimensionality, Eq. (1) cannot be solved exactly in practice. Powell (2011) describes four classes of approximation methods to obtain a policy: policy approximation, look-ahead, value function approximation, and cost function approximation. The last class involves using a parametric cost function to determine the policy as shown in Eq. (2), where $\bar{C}_t^\pi(S_t, x|\theta)$ is a parametrically ($\theta$) modified cost function subject to a parametrically modified set of constraints to account for the uncertainty.

$$X^\pi(S_t|\theta) = \arg \min_{x \in X_t(\theta)} \bar{C}_t^\pi(S_t, x|\theta) \quad (2)$$

The cost function can be further re-written as: $\bar{C}_t^\pi(S_t, x|\theta) = \bar{C}_{d,t}^\pi(S_t, x) + \theta \bar{C}_{r,t}^\pi(S_t, x)$ where $\bar{C}_{d,t}^\pi(S_t, x)$ is the delay cost and $\bar{C}_{r,t}^\pi(S_t, x)$ is the rebalancing cost. In the context of rebalancing idle mobility-on-demand vehicles, Sayarshad and Chow (2017) proposed a relocation policy as a cost function approximation, where the relocation cost function is modified to approximate the future costs in the system using M/M/s queue delay. We propose a policy that builds on this CFA methodology. The cost function is formulated as a mixed integer linear programming (MILP) problem with minimum cost flow relocation using the node-charge graph structure in the next section.

### 3.2. Rebalancing and charging cost function approximation policy
#### 3.2.1 Node-charge graph approach

We propose a method of solving the MDP problem using CFA as shown in Eq. (2) by specifying a relocation cost function within a new graph structure. We adopt a node-charge graph approach as illustrated in a one-dimensional network in Figure 2 without loss of generality.

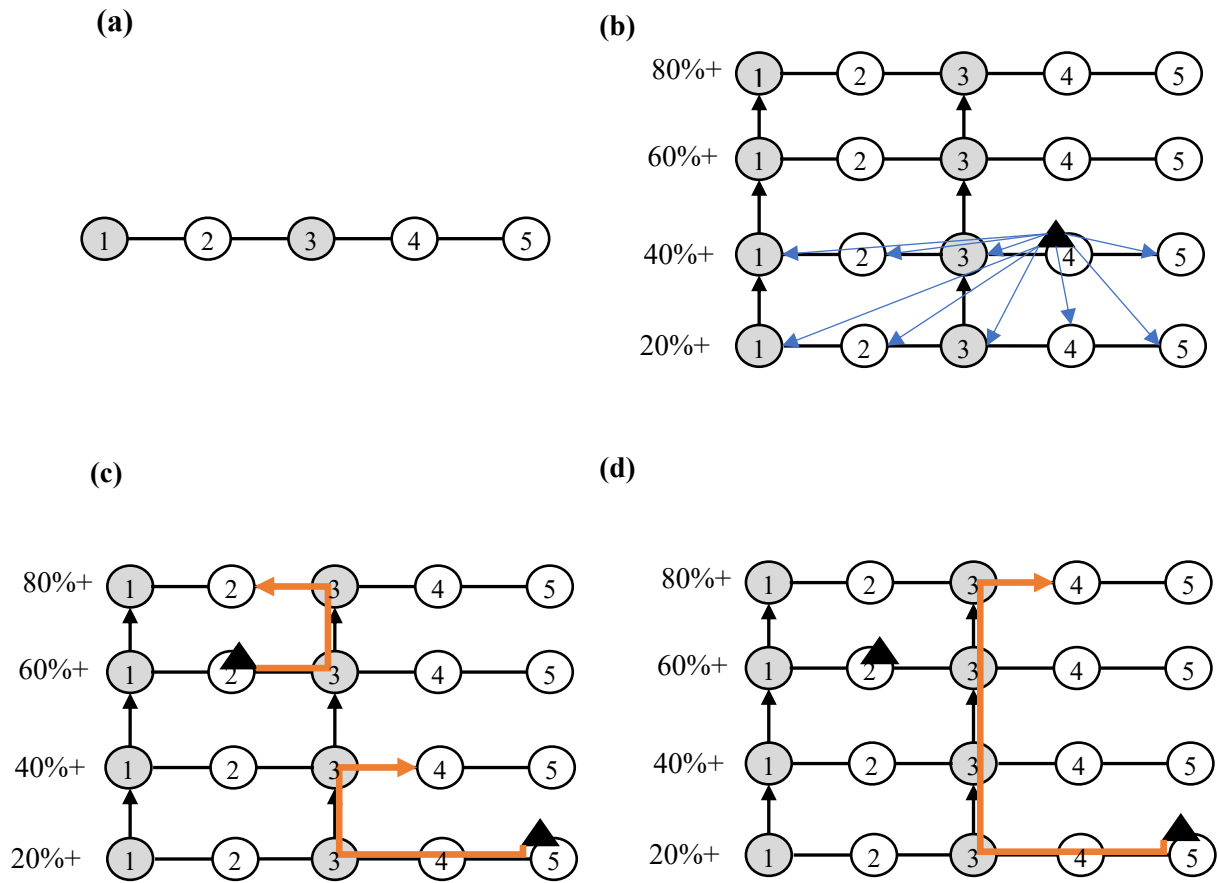

**Figure 2.** (a) initial graph; (b) expansion to a node-charge graph with coverage illustration in blue arrows; (c) one possible rebalancing solution in red arrow; and (d) a second rebalancing solution.

Consider a 5-node network lined up in a one-dimensional sequence as shown in Figure 2(a), where node 1 and node 3 are charging stations (denoted by gray nodes), i.e. $J = \{1,3\}$. The graph is expanded into a node-charge graph representation from $i \in N$ to $(i, h) \in (N, H)$, where $H$ is a set of discrete charge levels. Each layer represents the same zones at a certain charge level. Unidirectional links are added connecting each charge layer in node $i \in J$ going from one charge level up to the next higher charge level. These links have costs representing charging cost and time for the operator. Figure 2(b) illustrates the expanded graph, where each layer is a different charge interval (e.g. 20%+, 40%+, 60%+, 80%+). For example, a charge level of 20%+ refers to a charge of at least 20% and less than 40%.

A vehicle positioned at a node covers all nodes with lower charge as well. This ensures that a person seeking a vehicle with at least 40% charge would also be happy booking a vehicle with 60% charge. For example, a vehicle at node (4, 40%+) can serve nodes 1 to 5 at 40%+ and at 20%+, as illustrated by the blue arrows in Figure 2(b).

Access costs for demand are based only on the spatial link costs and not the charging link costs. The access cost of demand at (2, 20%+) for the vehicle at (4, 40%+) is just the cost from node 2 to node 4.

The charging stations are capacitated. This capacity is represented not by individual link capacities between charging layers within the node-charge graph, but by the sum of all path flows through each vertical column. For example, suppose there are two idle vehicles, one at (5, 20%+) and one at (2, 60%+). Two feasible rebalancing solutions are shown in Figure 2(c) and 2(d). In Figure 2(c), one vehicle is directed towards station 3 to recharge from 60% to 80% before being relocated back to node 2, while the second vehicle is sent to station 3 to recharge from 20% to 40% before being relocated to node 4. This solution requires a capacity of 2 vehicles at node 3. In Figure 2(d), only one vehicle is relocated to station 3 to be recharged from 20% to 80% and would then be relocated to node 4. This only requires a capacity of 1 vehicle at node 3. The two solutions lead to different coverage results and different charging capacity requirements. In addition, the relocation involves paths within the entry and exit charge levels for a given charge node, not simply direct flow from a surplus node to a sink node as was the case in prior relocation models, e.g. Chow and Sayarshad (2017), Nair and Miller-Hooks (2009), and Chow and Regan (2011). This implies a minimum cost flow problem with path-based bundle capacities at each charging station zone.

The optimality of these solutions depends on a mix of factors. For example, Figure 2(c) may be best if there is high demand near nodes 1, 2, and 3 up to 80%+ charge and there is enough capacity at station 3 to allow two vehicles to charge at the same time. The charging cost might be very high relative to the spatial relocation and/or access costs/penalties, leading to two short charges instead of one longer charge. Alternatively, Figure 2(d) may be best if the high charge demand is located closer to nodes 4 and 5, and/or perhaps there is only enough capacity for 1 vehicle charging at station 3 while the relocation cost to station 1 does not warrant the additional charging.

This graph structure can be set to different discrete charge levels. The number of levels determines the computational complexity of the problem. Each additional charge level duplicates the network. A model with $|H|$ charge levels has up to $\frac{(|H|)(|H|-1)}{2}$ different paths per charging station node to keep track of (for 10 levels that is 45 paths).

We assume that charged vehicles will not occupy a charging slot once fully charged. The bottom charge level can represent the minimum charge needed to get from any zone to any other zone without depleting fully.

### 3.2.2 Route-capacitated minimum cost flow relocation

**Parameters:**
$N$:     set of nodes
$H$:     set of charging levels, $h = 1,2, ..., |H|$
$A$:     set of directed arcs in the node-charge graph, $A = \{((i,g),(j,h))|\forall i, j \in N, g, h \in H\}$
$B$:     total number of idle vehicles at the start of a rebalancing time interval
$y_{ig}$:     number of idle vehicles at the node-charge $(i,g)$ at the start of a rebalancing time interval
$J \subset N$:   subset of nodes that are charging stations
$\theta$:     rebalancing cost parameter
$u_j, j \in J$: capacity of charging facilities available
$\tau_{ij}$:     user access cost from node $i$ to $j$
$\mu_{jh}$:     service rate parameter at node-charge $(j,h)$
$\rho_{\eta jhm}$:   utilization rate of the $m^{th}$ server (vehicle) with reliability threshold $\eta$ at node-charge $(j,h)$
$c_{igjh}$:   cost on an arc of node-charge graph from $(i,g)$ to $(j,h)$
$O$:     set of origins of idle vehicles on the node-charge graph
$\lambda_{ih}$:   arrival rate of customers at node $i$ with demand for $h$ charge levels or higher; assume that customers use up exactly $h$ intervals during their trip;
$C$:     max number of servers at a node-charge
$F$:     set of idle vehicles at the beginning of idle vehicle relocation epoch
$M$:     large positive penalty constant
$A^+_{ig}$:   set of outgoing arcs originating at a node-charge $(i,g), i \in N, g \in H$
$A^-_{ig}$:   set of incoming arcs destinated to a node-charge $(i,g), i \in N, g \in H$

**Decision variables:**
$W_{igjh}$:   rebalancing EV flow on arc $((i,g),(j,h)) \in A$
$Y_{jhm}$:   1 if the m-th vehicle are located at node $j$ with charge $h$, 0 otherwise
$X_{igjh}$:   vehicle at node $j$ with charge level $h$ serves customer demanding $g \in \{1, ..., |H|\}$ charge or more at node $i$ if $X_{igjh} = 1$, and 0 otherwise
$p_{jgjh}$:   path flows entering charging node $j \in J$ at charge level $g$ and exiting at level $h > g$

    Let $G'(N',A)$ be a directed graph with $N'$ being a set of node-charges $(i,h), \forall i \in N, h \in H$, and $A$ a set of directed arcs, $A = \{((i,g),(j,h))|\forall i,j \in N, \forall g, h \in H\}$. A node-charge $(i,h)$ is characterized by demand location $i \in N$ and requested minimum charging level (battery level) $h \in H$. All demand between two levels $h-1$ and $h$ sum up to level $h$. Arcs only cross from one charging level up to another at charging station nodes $J$ and there is a capacity $u_j > 0, j \in J$, applied to all flows through that station node regardless of charge level. Charging arcs belong to a subset of arcs defined as $A_\hbar = \{((j,g),(j,h))|\forall j \in J, \forall h = g+1, g = 1, ... |H-1|\}, A_\hbar \subset A$. The assigned flow on arcs is an integer decision variable. The cost of assigning flow on arcs is the

multiplication of arc flows by their unit costs, measured as rebalancing time/cost or charging time/cost on arcs.

The cost function is shown in Eq. (3) – (18). The parametric modifications corresponding to $\theta$ in Eq. (2) include the use of the arrival rates in Eq. (3) and constraint (7), whose parameters can be updated over time with new information from $S_t$ (e.g. new customer arrival rates, reservation length).

$$\min \bar{C}_t^\pi = \bar{C}_{d,t}^\pi(S_t, x) + \theta \bar{C}_{r,t}^\pi(S_t, x)$$
$$= \sum_{i \in N} \sum_{j \in N} \sum_{h \in H} \sum_{g \in H} \lambda_{ig} \tau_{ij} X_{igjh} + \theta \sum_{((i,g),(j,h)) \in A} c_{igjh} W_{igjh} \quad (3)$$

s.t.

$$\sum_{j \in N} \sum_{h \in H, h \geq g} X_{igjh} = 1, \quad \forall i \in N, g \in H \quad (4)$$

$$\sum_{j \in N} \sum_{h \in H, h < g} X_{igjh} = 0, \quad \forall i \in N, g \in H \quad (5)$$

$$Y_{jhm} \leq Y_{jh,m-1}, \quad \forall j \in N, h \in H, m = 2, 3, \dots, C \quad (6)$$

$$\sum_{i \in N} \sum_{g \in H} \lambda_{ig} X_{igjh} \leq \mu_{jh} \left[ Y_{jh1} \rho_{\eta jh1} + \sum_{m=2}^{C} Y_{jhm}(\rho_{\eta jhm} - \rho_{\eta jh,m-1}) \right], \forall j \in N, h \in H \quad (7)$$

$$\sum_{j \in N} \sum_{h \in H} \sum_{m=1}^{C} Y_{jhm} = B \quad (8)$$

$$X_{igjh} \leq Y_{jh1}, \quad \forall i, j \in N, g, h \in H \quad (9)$$

$$\sum_{(j,h) \in A_{ig}^-} W_{igjh} - \sum_{(j,h) \in A_{ig}^+} W_{igjh} \leq M Y_{ig1}, \forall (i, g) \in N \times H \backslash O \quad (10)$$

$$-\left( \sum_{(j,h) \in A_{ig}^-} W_{igjh} - \sum_{(j,h) \in A_{ig}^+} W_{igjh} \right) \leq M Y_{ig1}, \forall (i, g) \in N \times H \backslash O \quad (11)$$

$$\sum_{(i,g) \in A_{jh}^-} W_{igjh} - \sum_{(j,h) \in A_{ig}^+} W_{igjh} + y_{ig} = \sum_{m=1}^{C} Y_{jhm}, \quad \forall j \in N, h \in H \quad (12)$$

$$\sum_{g' \leq g} \sum_{h' \geq h} p_{jg'jh'} = W_{igjh} \, \forall j \in J, g \in H, \forall ((j,g),(j,h)) \in A \quad (13)$$

$$\sum_{g=1}^{|H|-1} \sum_{h'=g+1}^{H} p_{jgjh'} \leq u_j, \forall j \in J, g \in H, \forall((j,g),(j,h)) \in A \tag{14}$$

$$X_{igjh} \in \{0,1\}, \forall i,j \in N, g,h \in H \tag{15}$$

$$Y_{jhm} \in \{0,1\}, \forall j \in N, h \in H, m = 1,2,3,\ldots,C \tag{16}$$

$$W_{igjh} \geq 0, \forall i,j \in N, g,h \in H \tag{17}$$

$$p_{jgjh} \geq 0, \quad \forall j \in N, g,h \in H \tag{18}$$

The objective function minimizes the total access cost of customers (which can include generalized costs and costs of scheduling via reservations, etc.) to idle vehicles and total routing cost of idle vehicles (i.e. travel time/cost from the current locations of idle vehicles to charging stations, charging time/cost and travel time/cost to its respective destinations) on the node-charge graph. The rebalancing operations are run at each predefined time interval in order to serve customer demand and minimize queueing delay and operating cost as part of the online policy.

Constraints (4) and (5) require that rebalanced idle vehicles serving randomly arriving customers have sufficient charge. Constraint (6) is an order constraint stating a m-th server can be present only if there is already a (m-1)-th server at the same location.

Constraint (7) is the piecewise linear queueing constraint from Marianov and ReVelle (1996) representing the intensity requirement, where $\sum_{i \in N} \sum_{g \in H} \lambda_{ig} X_{igjh}$ is arrival rate and $\mu_{jh}$ is the service rate representing the amount of time a customer takes out a vehicle before returning it to the system. $[Y_{jh1}\rho_{\eta jh1} + \sum_{m=2}^{C} Y_{jhm}(\rho_{\eta jhm} - \rho_{\eta jh,m-1})]$ is a piecewise linear expression that captures the desired intensity for a given number of servers. For example, if there are $m$ servers, it is $\frac{\lambda}{\mu} = \rho_m$, where $\rho_m \leq m(1-\eta)$, where $\eta$ is the threshold desired (i.e. if $\eta = 0.1$ it means the model ensures intensity does not exceed 90% of maximum), and $\rho_m$ can be split recursively into contributions that each server adds. Given a user-defined intensity threshold $\eta$, $m$ idle vehicles (servers) and $b$ customers in a queue, the value of $\rho_{\eta jm}$ can be obtained exogenously by solving for the binding value of $\rho$ in Eq. (19), derived in Marianov and ReVelle (1996).

$$\sum_{k=0}^{m-1} ((m-k)m!\,m^b/k!)\,(1/\rho^{m+b+1-k}) \geq 1/(1-\eta) \tag{19}$$

Constraint (8) states the total number of servers is equal to the total number of available idle vehicles. Constraint (9) assures that only a location with servers can cover demand nodes. Since Eq. (4) allows for nodes to cover lower charge nodes, combining that with Eq. (9) implies that coverage extends to lower charge levels. Constraints (10) – (12) are the flow conservation constraints of the minimum cost flow problem.

For the charging station capacity, we need to ensure that the assigned flow on charging arcs do not exceed the limit of chargers available at a charging station. As mentioned earlier, this is not a link capacity but should be modeled as the sum of all path flows through any of the links corresponding to the charging node. This constraint should ensure that, for example, Figure 2(c) should not occur if $u_2 = 1$ because technically both vehicle flows are concurrently using that charging station. To address this, the link flows $W_{igjh}$ are matched to enumerated path flows in constraint (13). There is one set of path flows for each charging station, so there are not many – for 4 charging levels there are 6 variables per charging station: e.g. {1-2, 1-3, 1-4, 2-3, 2-4, 3-4} where charge level 1 is lower than charge level 2. The path flows are used to ensure that path flow capacity is met in constraint (14). This is a significant change in model formulation from prior models.

Lastly, $X_{igjh}$ and $Y_{jhm}$ are binary decision variables (Eq. (15) – (16)). Sayarshad and Chow (2017) showed that $Y_{jhm}$ can generally be relaxed to a continuous variable between [0,1] since the piecewise linear constraint will be satisfied, which leads to a much more computationally efficient model. Arc flow $W_{igjh}$ is a non-negative vehicle relocation flow (which will be integer due to the Unimodularity Theorem for the minimum cost flow problem), and the path flows $p_{jgjh}$ are continuous non-negative variables (Eq. (17) – (18)).

When Eq. (7) is relaxed, the model becomes a myopic case that does not anticipate the steady state delay in the system of a current decision. When $b \to \infty$ the chance constraint should allow for any queue length, and as $\eta \to 0$ any intensity is allowed.

As a p-median problem with additional constraints, the cost function is NP-complete (see Owen and Daskin, 1998). Existing heuristics for p-median problems like Teitz and Bart (1968) are not directly applicable because they violate queueing intensity and capacity feasibility. We propose a new heuristic to solve this problem so that it can be applied in an online setting for realistic fleet sizes of hundreds of vehicles like that of ReachNow in Brooklyn in 2017.

### 3.3. Proposed greedy heuristic

In order to provide a computationally efficient rebalancing system for large networks, we propose a heuristic algorithm for solving the cost function as part of the CFA policy. A rebalancing system should scale up to a fleet like the BMW ReachNow one in Brooklyn, NY. This target network has 303 zone centroids and considers up to 5 charge levels.

Computational tests on a range of random instances with up to $|N| = 1000$ and $|H| = 4$ are solved using exact algorithms from commercial software (MATLAB), but run times can exceed 2 hours (see Section 4.2) with an Intel i5-6300U CPU with 2 cores and 8GB memory. Clearly this would not be feasible to run in an online setting with commercial software.

We propose a heuristic that solves such instances more computationally efficiently. The core of the algorithm is based on the greedy heuristic from Teitz and Bart (1968) but modified to maintain feasibility with respect to: (a) queueing constraints, (b) capacity, (c) minimum cost flow relocation, and (d) demand access savings when accounting for multiple servers. In a p-median problem without queueing constraints, each server can satisfy the entire demand from any location. In our model, however, a server cannot satisfy demand at higher layers, while demand at layers lower or equal to the server's location can be served up to the RHS amount of Eq. (7). For this reason, a node that has an idle car already can still yield potential gains by adding an additional server to it. A summary is provided in Algorithm 1, which is designed to run at the start of each discrete time interval in an online system as a CFA policy.

**New variables definition**
$k$: iteration step
$S^k$: savings matrix in iteration k
$N^k$: set of infeasible relocation points in iteration k

$A^k$: minimum access cost matrix in iteration k
$L_f^k$: location of vehicle $f$ in iteration k
$d_1$: Total demand assigned to the idle vehicle at $(j,h)$ during the previous iteration
$d_2$: Total demand that can be assigned to the idle vehicle at $(j,h)$
$e$: number of vehicles that violated charging capacity

**Algorithm 1: Proposed greedy heuristic**

1. **Initialize parameters:** $k = 0$, $N^k = \emptyset$, $e = 0$, $0_{S_{|N|,|H|}}$, $0_{A_{|N|,|H|}}$, $B = |F|$
2. **While** $k \leq B$ **do**
3.     **For** every $s_{i,g} \in S^k$
4.         **If** $(i,g) \in N^k$ or $g^{k,k} < g$
5.             $s_{i,g} := -\infty$
6.         **Else If** $Y_{jh1} = 0$
7.             $s_{i,g} := \sum_{j \in N} \sum_{h \in H} [\tau_{ij} - a_{j,h}]^+ \lambda_{jh} - \theta c_{igjh}$, where $a_{j,h} \in A^{k-1}$
8.         **Else**
9.             $d_1 := \sum_{j \in N, h \in H} \lambda_{jg} X_{igjh}$,
10.            $d_2 := \min\{d_1 - \rho_{jh,m} \mu_{jh}, (\rho_{jh,m+1} - \rho_{jh,m}) \mu_{jh}\}$,
11.            $s_{i,g} := \frac{d_1}{d_2} s'_{i,g}$, where $s'_{i,g} \in S^{k-1}$
12.         **End If**
13.     **End For**
14.     **If** $k = 0$
15.         $L_f := \arg \max_{i \in N, g \in G} (S^k)$, $\forall f \in F$
16.     **Else**
17.         $N^M := N^M \cup \{(i,g) | Eq.(7) \text{ not satisfied}\}$
18.         $L_k := \arg \max_{i \in N, g \in G'} (S^k)$, where $G' = \begin{cases} g_0, & \text{if } e > 0 \\ H, & \text{otherwise} \end{cases}$
19.     **End If**
20.     $e := \sum_{f \in F} sgn(g^{k,f} - g^{0,f}) - \sum_{j \in J} u_j$
21.     **Update** $A^k$: $a_{j,h} = \{\min t_{ij} X_{igjh} | \sum_{j \in N} \sum_{h \in H, h \geq g} X_{igjh} = 1, \sum_{j \in N} \sum_{h \in H, h < g} X_{igjh} = 0, (i,g) \in L_k\}$
22.     $k = k + 1$
23. **End While**

The algorithm is initiated with a set of vehicle locations: $L = (L_f)_{f \in F}$, where each location $L_f$ is expressed by the node-charge tuple $(i_f^0, g_f^0) \in (N, H)$. The heuristic computes a savings matrix $S^k$ at each iteration k (**steps 3 – 13**) and makes rebalancing decisions accordingly. The accessibility matrix $A^k$ (**step 21**) stores the closest distance to an idle vehicle for each network node $j \in N, h \in H$, but not all nodes may be assigned to their closest idle vehicle due to the feasibility constraint (7).

To address this limitation, we calculate the feasible quantity $d_2$ that can be assigned at each node (**step 10**). The fraction of quantity unserved $d_1/d_2$ is used to update the savings using information from the previous iteration. During the first iteration of the algorithm, given an initial vehicle assignment $L_0$, all vehicles are placed at a single location (**step 15**) that minimizes total access costs. Since $a_{j,h} = 0 \ \forall j, h$, the savings minimization objective is equivalent to Eq. (3).

The reasoning behind placing all idle vehicles at a single location is to ensure that the solution remains feasible under constraint (7). To ensure that subsequent relocation decisions remain feasible, candidate nodes that violate Eq. (7) are included in set $N^k$ (**step 17**) and are not considered for relocation (**step 18**). In the case that the initial solution is infeasible, we can conclude that the MILP is also infeasible. For discrete time intervals where the MILP is infeasible, constraint (7) is relaxed to solve a myopic problem in those cases (which is always feasible).

Since this problem includes a capacity constraint Eq. (14), we calculate the amount of excess capacity (**step 20**) used to make the relocation decision $L_K$. For all subsequent relocation decisions (**step 18**), we consider only a subset of nodes $G' = g_0$ until constraint (14) also becomes feasible ($e = 0$). The node-vehicle assignment is performed every time a vehicle location is updated (**step 21**). The process ends (**step 23**) when the iterative procedure has been performed for all $B$ idle vehicles. The complexity of the algorithm is $O(N^2|F|)$.

In situations where non-myopic rebalancing is infeasible, the optimization step is repeated with relaxed queuing constraint requirements for the CFA policy.

## 4. Model and heuristic analysis

### 4.1. Cost function verification test

Evaluation of the proposed cost function and greedy heuristic are conducted in two sets of replicable experiments. The first is with an illustrative example that serves to verify the mixed integer programming model formulation and demonstrate the capabilities to evaluate certain trade-offs.

Consider a small network of 24 node-charge nodes corresponding to 6 original zones (labeled from 1 to 24 for convenience) extended up to 4 charging demand levels (20%, 40%, 60%, 80%) shown in Figure 3. The goal is to test whether the proposed model can effectively rebalance the idle vehicles to meet all customer demand under available charging capacity constraints. The travel time between nodes is denoted as $\tau_{ij}$ and the charging time (vertical travel distance) as $c_{ij}$. Three idle vehicles with respective remaining charge levels are located at node-charge 3, 7, and 14. We consider 2 charging stations (nodes 2 and 6) with the same capacity $u$ per station. We test the model under three different capacities: $u = \{1,2,3\}$. Customer arrival rates are arbitrarily generated over all node-charges and fixed for the three scenarios. These rates are shown in the numbers over each node-charge (e.g. 3.8 customers/hr arrive at node-charge 15 corresponding to node 3 at charge level 60%+). Demand varies between different charging levels because the range required for specific trips can differ among travelers.

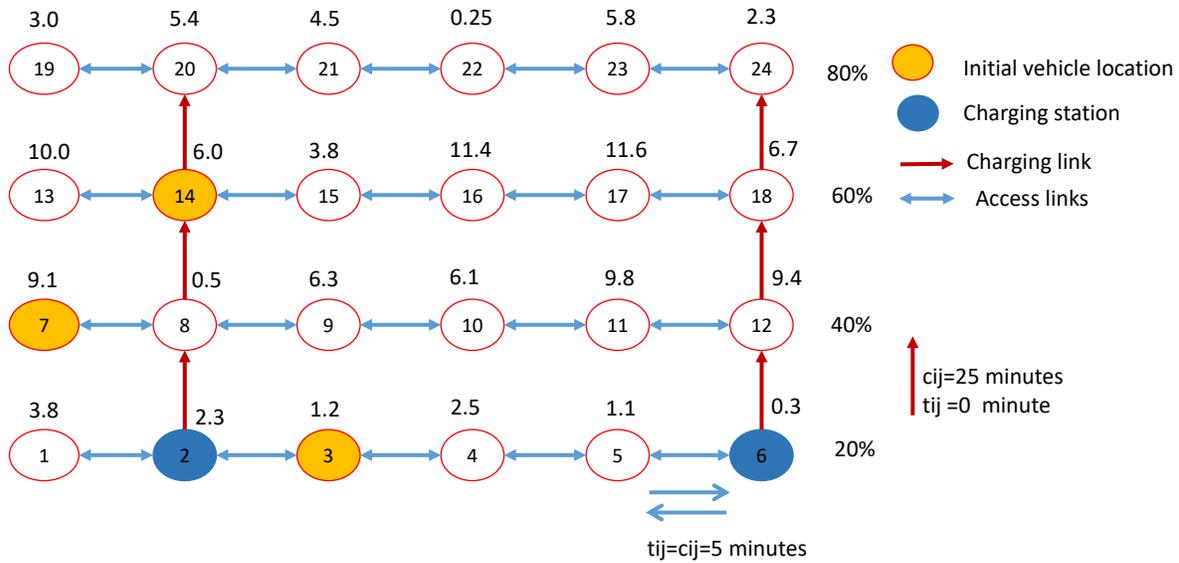

**Figure 3.** Test instance used to evaluate model.

This test is implemented using a Dell Latitude E5470 laptop with win64 OS, Intel i5-6300U CPU, 2 Cores and 8GB memory. The MATLAB mixed-integer linear programming solver (intlinprog) is used to obtain exact solutions. The test instances are publicly available on Zenodo (Chow, 2019).

Figure 4 presents the computational result of rebalancing EV flows for the case of capacity $u = 3$. We see that all vehicles are rebalanced to the highest level 4 at different nodes which minimize total access cost of customers. The vehicles are assigned to use the nearest charging links to their destinations. The optimal objective value is $Z^* = 281.5$.

The results for the other two capacity scenarios are shown in Figure 5. When reducing the capacity by 1 unit to $u = 2$, the first charging station at node 2 becomes fully capacitated by vehicles at node 7 and 14. The vehicle at node 3 moves farther to charge at node 6 and comes up to its destination at node 23, as shown in Figure 5(a). The numbers over the links represent the assigned flow on links. The obtained $Z^* = 281.5$ is the same as the preceding case since the relocation costs are the same as the prior case while the final assigned locations are identical.

When further reducing the capacity to $u = 1$ per station, there are only two chargers available in the system. The solution shows the vehicles at node 7 and node 3 coming up to charge level 80% as before. However, the vehicle at node 14 is rebalanced on the same level to node 16 without recharging further due to the limited capacity. All charging capacity is used with least system cost as shown in Figure 5(b). The obtained objective values is now increased to $Z^* = 300.25$. These results verify the model formulation and the trade-offs that it can analyze.

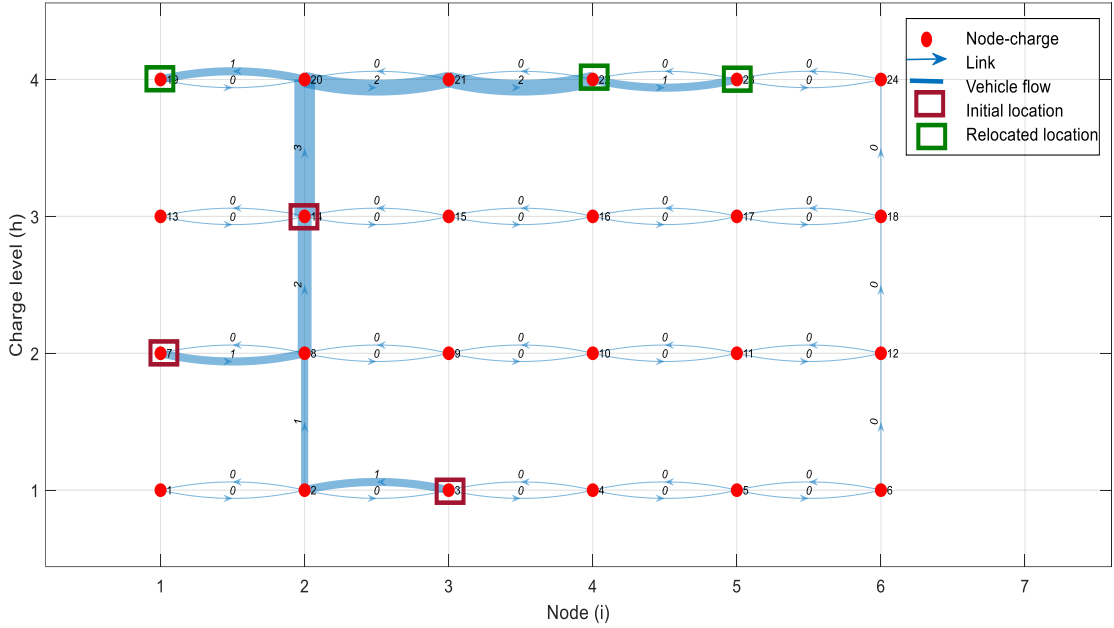

**Figure 4.** Rebalancing idle vehicle flows and its locations under charging station capacity $u = 3$.

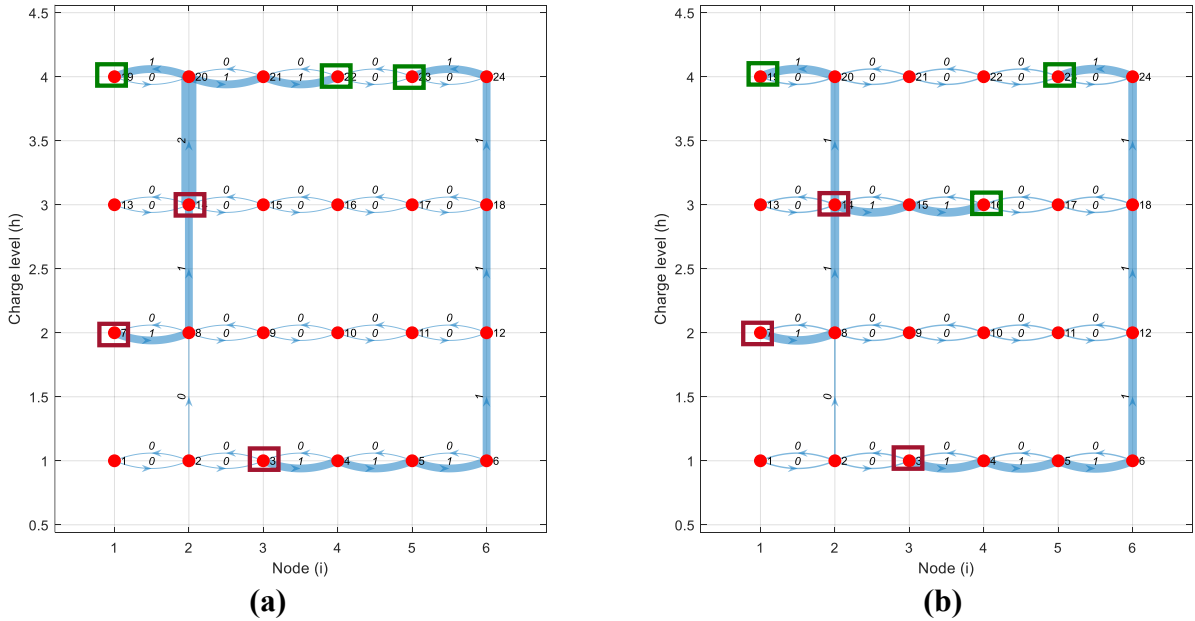

**(a)**  **(b)**

**Figure 5.** Rebalancing idle vehicle flows under charging station capacity (a) $u = 2$ and (b) $u = 1$.

## 4.2. Model scalability tests on proposed heuristic

A second set of experiments is conducted with a set of 5 generated instances ranging in size from 10 to 200 nodes, and 5 charging levels (up to 1,000 node-charge nodes) to test the scalability of the model using commercial solvers and the proposed heuristic. More charging levels |H| lead to increased accuracy in assigning idle vehicles to appropriate node-charges, but with a trade-off in computational cost since paths need to be enumerated along the charge levels for a given node. With 5 charging levels, it assumes customers differentiate between 50-75% or 25-50% charge levels, and when they take a vehicle out, they may drive it in increments of 35 miles or so (for a 140-mile range vehicle). The tests in this section and subsequent ones were implemented on MATLAB 2017a with a MacBook Air 2.2 GHZ i-7 8GB RAM with 1600 MHZ DDR3 OS Catalina 10.15.4.

Customer demand at each node-charge is randomly generated between [0, 1]. The vehicle fleet |F| and number of charging stations are set to increase proportionally with the network size. The service parameter $\mu$ is set to represent medium arrival intensity so that constraint (7) will be binding. The parameter setting is shown in Table 1.

The results of the computational experiments are shown in Table 2. The MIP solution times increase exponentially with the number of nodes. The MIP performance can be verified in Ma et al. (2019b).

In online operations, computational savings are important and the heuristic results for networks up to 1000 node-charges are particularly promising. The optimality gap falls between 7 – 35% while the computational time is reduced by 15 – 89% compared to commercial solvers. These results suggest that the algorithm is suitable for deployment as an online rebalancing system. All generated instances can be found in Zenodo (Chow, 2019).

**Table 1.** Summary of parameter settings

| Parameter | Value |
|---|---|
| |N| | 10-200 |
| |H| | 5 |
| B | 0.4|N| |
| $\lambda_{ig}$ | Random number drawn from [0,1] for each node-charge |
| $\theta$ | 0.02 |
| C | 3 |
| $\rho_{jm}$ | (0.2236, 0.6416, 1.1576, 1.7345) |
| |J| | 0.1|N| |
| $u_j$ | |F|/(0.1|N|) |
| $\mu$ | $1.5 \times \sum_{i \in N} \sum_{g \in H} \lambda_{ig}$ |

**Table 2.** Computational times and optimality gap of the proposed heuristic over generated instances

| N | H | # of nodes ($N \times H$) | MIP comp. time (sec) | Optimality Gap (%) | Heuristic comp. time (sec) (% reduction) |
|---|---|---|---|---|---|
| 10 | 5 | 50 | 0.96 | 7.61 | 0.11 (-89%) |
| 20 | 5 | 100 | 0.46 | 9.63 | 0.19 (-59%) |
| 50 | 5 | 250 | 3.8 | 15.45 | 1.58 (-58%) |
| 100 | 5 | 500 | 16.01 | 18.49 | 10.63 (-34%) |
| 200 | 5 | 1000 | 144.17 | 34.77 | 122.78 (-15%) |

### 4.3. Additional sensitivity experiments

To further evaluate our heuristic, we perform sensitivity analysis on four different model parameters. For each parameter value, we generate one random instance. Our base-case scenario is the 10-node instance of Table 1 ($N = 10, H = 5$) with service rate $\mu = 30.47$, objective relocation cost factor $\theta = 0.02$, network demand of 20 passengers/hour ($\sum_{h \in H} \sum_{j \in J} \lambda_{jh} = 20.313$), and $B = 4$ idle vehicles. The results are summarized in Figure 6.

Larger vehicle fleets (Figure 6(b)) or additional charging stations (Figure 6(a)) do not have a major impact in solution quality. The model is infeasible if the number of idle vehicles is less than 3 because of the queueing constraint. A potential shortfall of the heuristic is revealed when service rate becomes stricter, represented in Figure 6(c) as a ratio of service rate over sum of arrival rates. The increased optimality gap suggests that the tighter queueing constraint bound makes it harder for the heuristic to match. When the multiplier is below 1.2 the model becomes infeasible. This is because the system becomes highly saturated such that the steady state assumption of Eq. (7) is no longer satisfied. Note that Figure 6(c) shows the service rate in declining value from left to right. The results illustrate that Algorithm 1 is sensitive to the service rate.

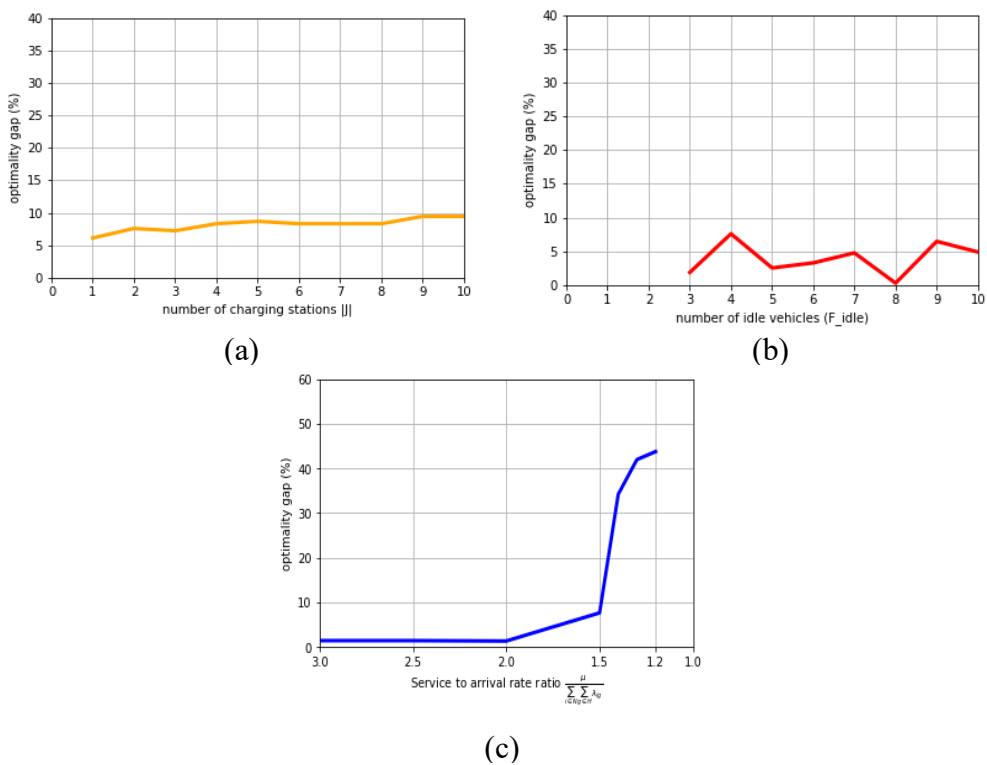

**Figure 6.** Heuristic sensitivity to (a) number of charging stations, (b) number of idle vehicle and **(c)** service rate to arrival rate ratio.

## 5. Simulation-based computational experiments

Having verified the cost function and algorithm in a single period, they are further tested as a CFA-based MDP policy in an online setting using simulation of customer arrivals, vehicle bookings, pickup and return time and locations. Within the simulation of this online environment, the rebalancing algorithm is run each time interval to determine rebalancing decisions. Both EV and non-EV scenarios are evaluated. For the EV scenarios, locations and numbers of electric charging stations are provided exogenously. Two sets of experiments are conducted: the first involves online simulation of the same small test network from Figure 3, while a second, large-scale instance is drawn from real customer arrival data and fleet size in Brooklyn, NY, to validate the algorithm. The real data was provided by BMW ReachNow.

Tests are conducted to validate the performance of the algorithm for the non-myopic case compared to a myopic setting as well as a benchmark operating policy. Measures of effectiveness include customer wait time, rebalancing costs, and number of customers in queue for a vehicle.

## 5.1. Simulation platform for evaluation

A custom event-based simulation platform with discrete rebalancing intervals was developed to evaluate different online system instances, as detailed in Li et al. (2019) and highlighted here. The simulator was developed in MATLAB. It is agent-based and includes two kinds of agents: customer agents and vehicle agents. The vehicle-to-customer assignment rules determine how the customers book the vehicles (in other words, how the vehicles are assigned to customers), and the rebalancing strategies determine how vehicles are rebalanced among different nodes. All the customer agents and vehicle agents operate in quasi-continuous time (unit time of 1 second) except for rebalancing decisions, which are made at the start of each (rebalancing) time interval. The event-based simulation with interval-based rebalancing decisions is illustrated in Figure 7, where vehicles may be simulated continuously over time in service but every time interval (t, t+1, …) the system may direct the idle vehicles to be rebalanced.

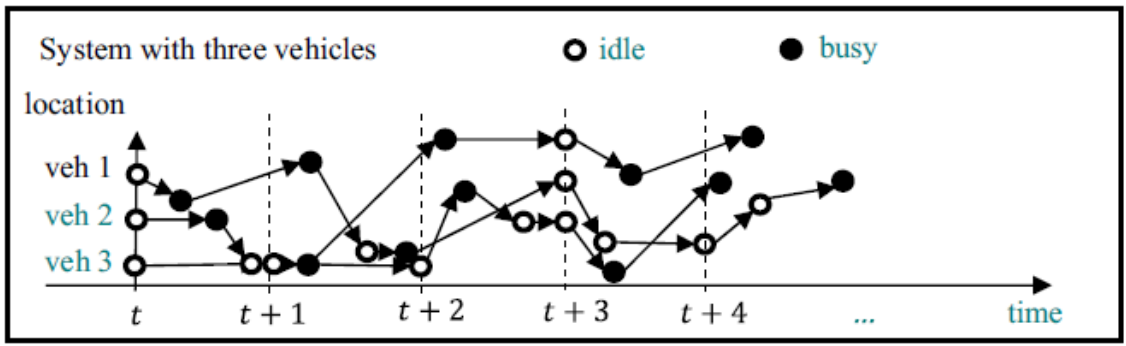

**Figure 7.** Illustration of discrete vehicle events over time overlaid with discrete time steps for rebalancing decisions (source: Sayarshad and Chow, 2017).

The rebalancing strategy is used to determine how to rebalance vehicles among different nodes. A rebalanced vehicle cannot be booked until it reaches its destination and changes its state to idle. The assignment rule ensures that customers cannot book a vehicle that is outside the pick-up distance and set to allow vehicles anywhere in the network to be rebalanced to any other part of the network. We can test different rebalancing strategies in the same simulation.

The parameter inputs depend on the rebalancing strategy. For the proposed policy, we need to feed the simulator the average arrival rate and service rate for each node, the location and charge level of all available vehicles, the driving distance and driving time between each pair of nodes, the location of all charging stations, and the number of chargers at each charging station. The output should specify the rebalancing route for all available vehicles, which is a sequence of nodes that the rebalanced vehicles should visit. If a vehicle is not assigned for rebalancing, then its route is just its current node.

The simulator is publicly available in Zenodo (Chow, 2019). A screenshot of the simulator output (for the Brooklyn study area) is shown in Figure 8. The + is the position of the traffic analysis zones (TAZs), while the size of the red square represents the number of available vehicles within this TAZ at an arbitrary time interval of $t = 820$.

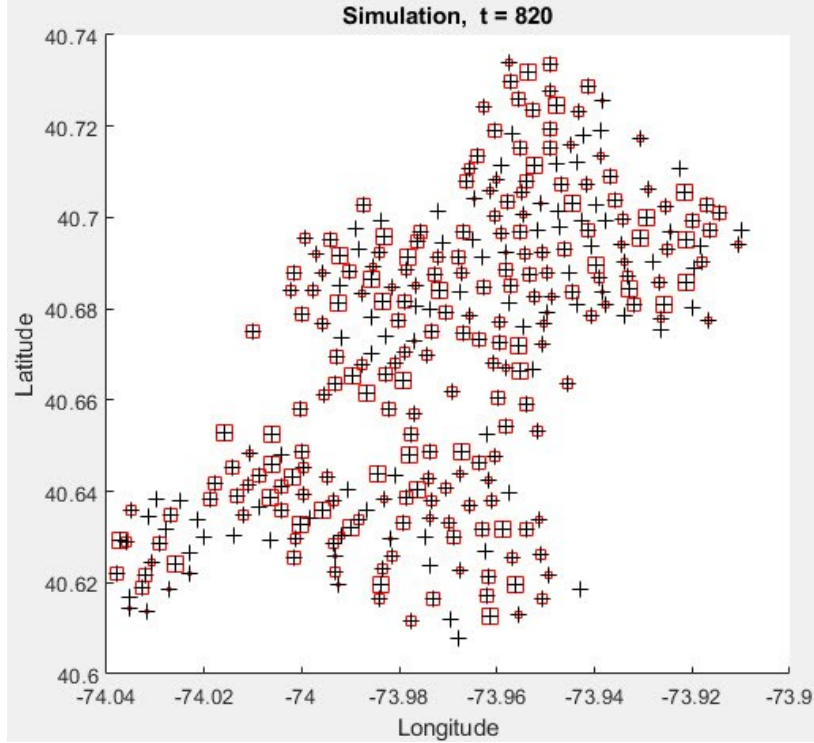
**Figure 8.** A sample screenshot of the simulator for Brooklyn, NY.

### 5.2. Small test instance: exact solutions

Using the simulator, we generated customer arrivals for the network shown in Figure 3. For the small network we set the rebalancing time step to be $\Delta t = 1\ minute$. Details of the simulation input parameters are shown in Table 3.

**Table 3.** Input parameters for small instance experiment

| Parameters | Value |
| --- | --- |
| Number of Nodes | 6 |
| Layers | 5 |
| Fleet | 20 |
| Station Node Locations | [2,6] |
| $\lambda_{ig}$ | Random number drawn from [0,0.1] for each node-charge |
| Station Capacities | 8 |
| Service constraint $\mu_{jh}$ (non-myopic case) | $5\ bookings/hour$ |
| Vehicle Average Speed | $60\ kph$ |
| Simulation duration | $5000\ minutes$ |
| Rebalancing time step interval $\Delta t$ | $1\ minute$ |
| Number of runs | 10 |

The simulation results shown in Figure 9 are promising. The piecewise linear queueing constraint manages to reduce waiting times by 22% when $\eta = 0.85$ and $b = 2$. The non-myopic strategy places idle vehicles strategically in anticipation of future demand. However, the largest improvement is observed when applying our algorithm to gasoline-fueled ("regular") vehicles. In this scenario both rebalancing and waiting times are significantly reduced compared to electric vehicles. This difference is attributed to the capacity and time constraints of electric car batteries.

Without rebalancing, not even regular vehicles perform adequately. This is shown in Figures 9(a) and 9(b), where the case without rebalancing has significantly higher waiting times and customers in the queue than any other scenario.

(a)

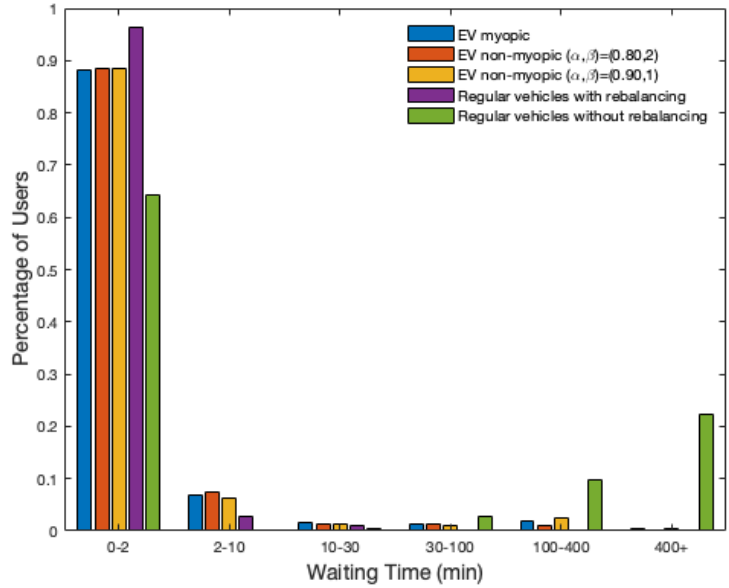

(b)

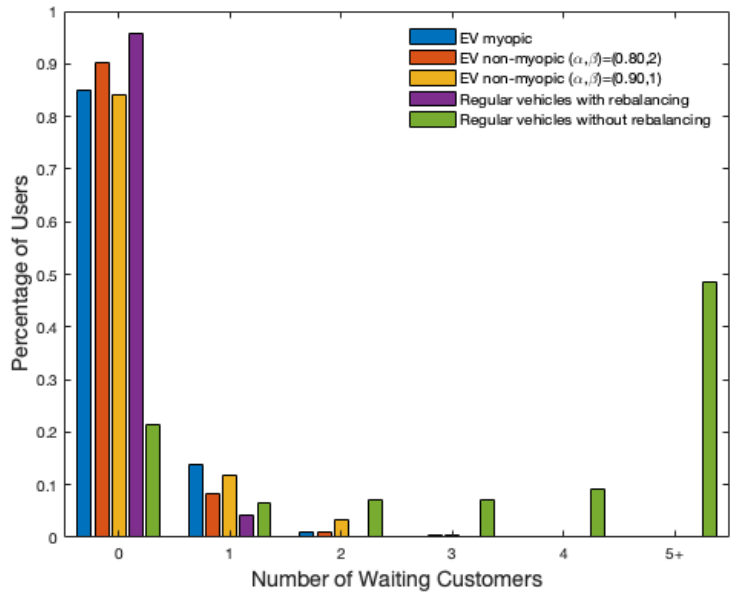

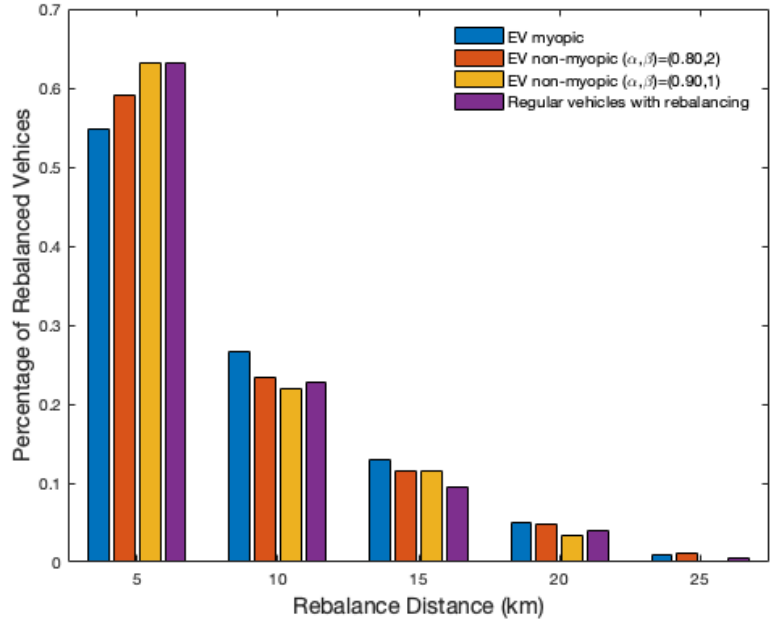

(c)

**Figure 9.** (a) Experienced waiting times, (b) customer queue lengths, (c) rebalance distance per period, averaged from 10 runs of each scenario.

Table 4 summarizes the statistics across all the scenarios. Each simulated scenario has a duration of 5000 minutes and the results are averaged across 10 runs. The summary quantifies the differences between electric fleets and regular vehicles. The calibration of parameters $\eta$ and $b$ are important in improving the solution quality of the non-myopic model and provide additional levers for operators to improve serve quality. In this experiment, we do not include a non-myopic rebalancing strategy for regular vehicles since it was already tested in Sayarshad and Chow (2017) and shown to reduce total costs by 27%. The test was further replicated in a multimodal setting in Ma et al. (2019a).

**Table 4.** Summary of statistics performance for simulation scenarios

| Scenarios: | EV *myopic* | EV *non-myopic* $\eta = 0.80, b = 2$ | EV *non-myopic* $\eta = 0.90, b = 1$ | Non-EV *with myopic rebalancing* | Non-EV *without rebalancing* |
|---|---|---|---|---|---|
| Waiting times *Mean* (min) | 8.289 | 6.445 | 10.690 | 2.215 | 279.231 |
| Waiting times *Std. dev.* (min) | 37.460 | 25.335 | 43.359 | 1.176 | 508.041 |
| Customers in queue *Mean* | 0.162 | 0.121 | 0.201 | 0.044 | 6.097 |
| Customers in queue *Std. dev.* | 0.395 | 0.372 | 0.469 | 0.208 | 4.801 |
| Rebalance distance *Mean* (km) | 8.532 | 8.271 | 7.765 | 7.331 | - |
| Rebalance distance *Std. dev.* (km) | 4.652 | 4.691 | 4.138 | 4.586 | - |

Algorithm 1 is used in place of the exact algorithm to obtain the cost function approximation policy for the same example. The comparison is shown in Table 5. The simulation results illustrate that the heuristic has a 58% optimality gap with respect to the realized cumulative costs corresponding to the objective (3) while reducing computation time by 46%. The computational savings that can be achieved in large networks can significantly offset the solution quality. The results that are shown in Tables 4 and 5 represent the average values of 10 simulation runs.

Table 5. Comparison of heuristic-based MDP vs MILP-based MDP

| Performance Metrics | MILP | Heuristic (% change in mean) |
|---|---|---|
| Waiting times (min) *[mean, median, std]* | [6.44, 2, 25.33] | [10.22, 2, 27.08] (+59%) |
| Rebalance distance (km) *[mean, median, std]* | [8.27, 5, 4.69] | [11.56, 10, 5.55] (+40%) |
| Customers in queue *[mean, median, std]* | [0.121, 0, 0.372] | [0.326, 0.1, 0.494] (+169%) |
| Computation time (sec) | 671.06 | 361.23 (-46%) |
| Average total cost (Delay + θ*RebDis) | 6.61 | 10.45 (+58%) |

### 5.3. BMW ReachNow case study data

We tested the algorithm in the Brooklyn network with the average monthly demand per TAZ shown in Figure 10 for 303 TAZs. We use data obtained from the BMW ReachNow car-sharing operations in 2017 to simulate passenger arrivals and destinations. Our dataset includes all trips for the month of September with an average of 231 car pickups per day. The service duration tends to have a much heavier tail than an exponential distribution; for example, the average reservation time is $\bar{x} = 6.5$ hours while the median is only $\tilde{x} = 45$ mins, as shown in Figure 11. Due to this discrepancy, we compute $\mu$ from the median $\left(\mu = \frac{\ln 2}{\tilde{x}}\right)$ instead of the mean, which is approximately 0.92 /hr. Although the median distance traveled per reservation is 4.1 mi, the average distance is 38.5 mi suggesting the vehicles are used to make several errands before being returned (contrary to the widely held assumption in prior studies that carshare vehicles are simply driven from pickup to drop-off).

As this data shows, even free-floating carsharing clearly should not be modeled as direct trips from a pickup location to a drop-off location. From the operator's side, the per-booking net revenue collected from operating the system is $22.60, while the after-tax cost for the average customer rises to $27.20.

To test the heuristic, we assume the fleet of 262 vehicles are electric with 18 charging stations and having an average capacity of 4 charging ports, located as shown in Figure 12. The locations of these stations are based on Chargehub (2019). The charging facilities already exist to serve privately owned electric vehicles. We run the simulation over a period of one month with rebalancing decisions made at hourly intervals. We test 8 different scenarios (*a – h*) and compare them using the metrics defined in the small case simulation (waiting time per customer, rebalance distances, etc.).

a) Myopic rebalancing for electric vehicles by using our model with the constraint (7) relaxed;
b) Proposed non-myopic rebalancing strategy using the proposed heuristic;
c) The proposed strategy with the heuristic operating under the current programmed prototype software in simulated real time;
d) The existing flee-floating BMW car-sharing data with regular vehicles that does not use any rebalancing algorithm;
e) A naïve "ChargerChasing" scenario that assigns vehicles to the closest charging facility right after they drop off a passenger, while giving priority to lowest charged vehicles to use charging ports;
f) Unlimited capacity for the initial 18 charging facilities as shown in Figure 10(a);
g) The addition of 5 charging stations with capacity of 4 ports per facility shown in Figure 10(b);
h) Non-EV with myopic rebalancing.

In practice, the algorithm is not solved instantaneously, so vehicles may end up not receiving rebalancing commands immediately at the start of a time interval. As a result, lag can occur between the rebalance decision and real time locations and availability of the vehicles. In order to examine the real-time efficiency of the proposed heuristic under the current programming language and software/hardware setting, we run an alternative version of the simulation (scenario c) to account for customer arrivals during the heuristic runtime. Instead of treating the optimization time as instant (see Figure 7), we allow vehicles to be picked up by customers while the algorithm is running. If a vehicle was assigned to be rebalanced upon algorithm completion but it is no longer available due to being picked up in real time, then it is skipped by the system. The other seven scenarios assume the algorithm is run near instantaneously, representing an ideal setting where the algorithm is implemented with more efficient programming language and hardware.

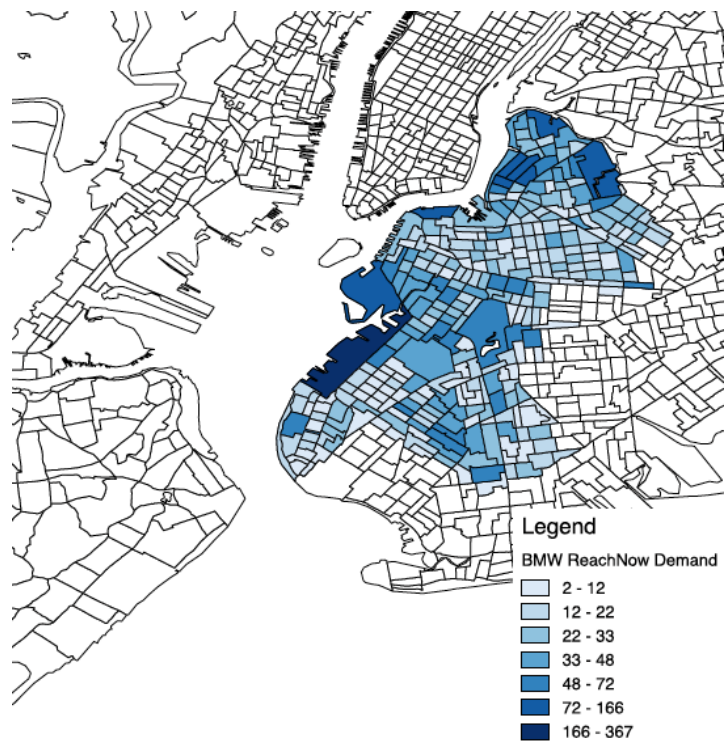

**Figure 10.** Brooklyn traffic analysis zones (NYMTC, 2010) with BMW ReachNow monthly demand.

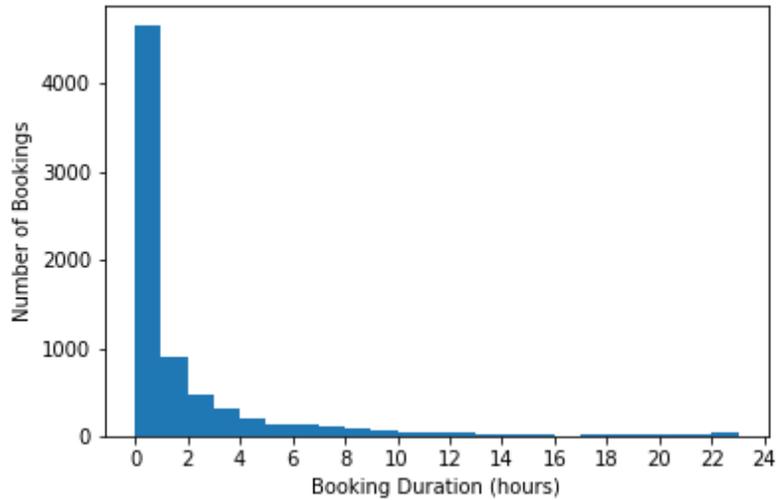

**Figure 11**. Vehicle booking duration histogram

All facilities are assumed to use fast DC chargers where vehicles can fully recharge in 30 minutes. A rebalancing cost parameter of $\theta = 0.02$ is assumed along with relocation, access, and charging link costs measured in same units of minutes. The simulation horizon is set to 30 days. Simulation time for the event-based simulation is rounded to the nearest minute. Time steps for rebalancing are set to $\Delta t = 1\ h$. The fleet size is 262 vehicles. There are 5 charge levels assumed. The other parameters used in each scenario are listed in Table 6.

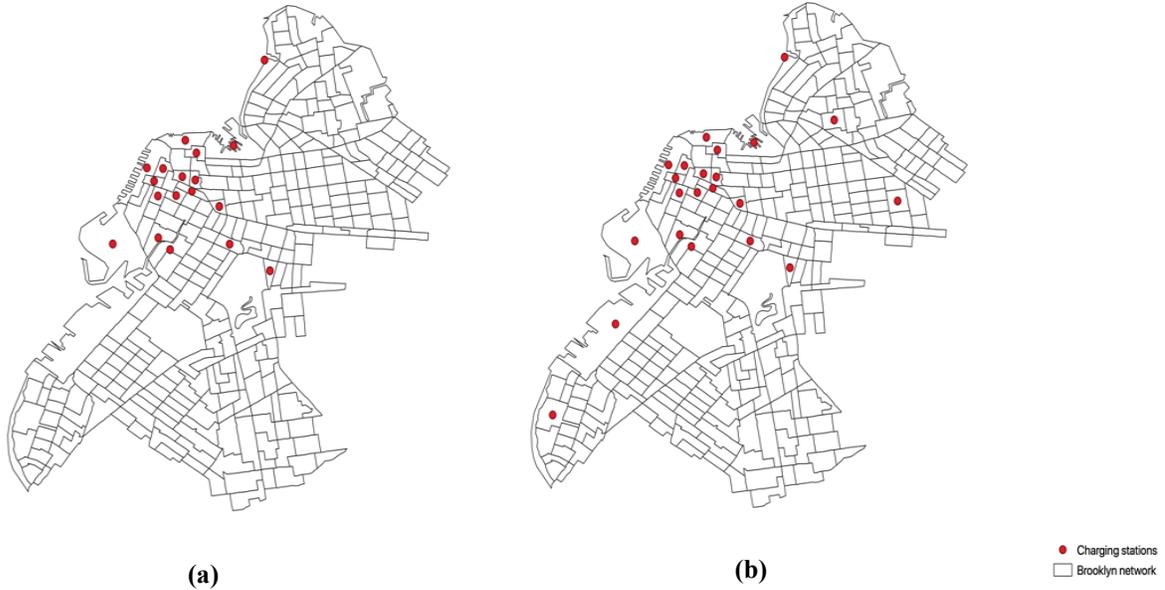

(a)  (b)

**Figure 12.** (a) Existing charging stations and (b) expansion scenario for EV fleet analysis.

The simulation step in Table 6 denotes the most granular level of time to track changes in the state of vehicle and passenger arrivals within the simulation environment. The vehicle range determines how far passengers can travel with a vehicle before refueling and the maximum charge level informs us about the number of clusters that battery percentage is split into. Finally, charge time denotes the amount of time it takes for a vehicle to charge from 0% up to 100%. In order to model regular fueled vehicles within the same simulation environment, we assume that they have a charging time of 1 minute (time it takes in gas station) and that every node is a charging station with infinite capacity, thus permitting vehicles to satisfy any demand cluster. Figure 13 illustrates the average number of idle vehicles for the current setting of ReachNow operations which do not include rebalancing. The heatmap is consistent with demand arrival patterns shown in Figure 10.

Table 6. Parameters used in the seven scenarios for the Brooklyn instance

| Scenarios | EV_Myopic | EV Non-Myopic | Non-EV No-rebalance | Charger-Chasing | EV Infinite Capacity | Add Charging Station | Non-EV Myopic |
|---|---|---|---|---|---|---|---|
| Vehicle range | 200 km | 200 km | 600 km | 200 km | 200 km | 200 km | 600 km |
| Charging station capacity | 4 | 4 | ∞ | 4 | ∞ | 4 | ∞ |
| No. of charging stations | 18 | 18 | 303 | 18 | 18 | 23 | 303 |

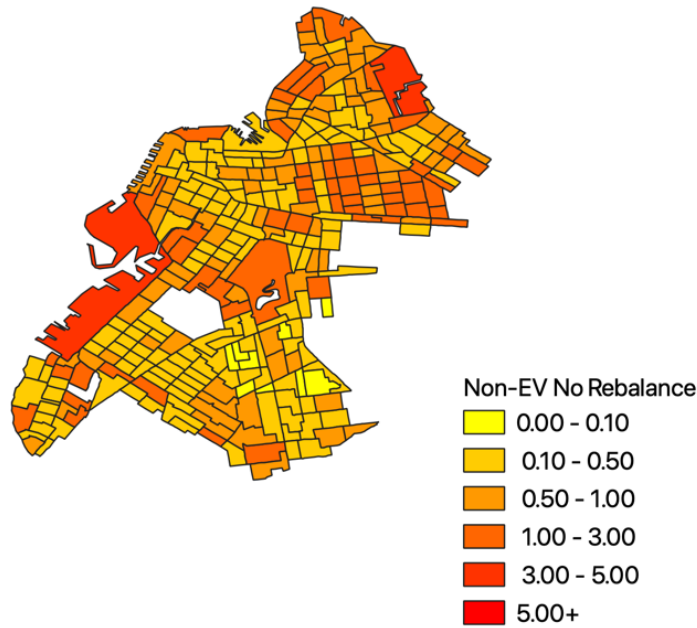

Figure 13. Idle vehicle heatmap of current operations without rebalancing

## 5.4. Brooklyn case study results

For these instances, we compare the rebalancing kilometers, average waiting time per customer, total delay and customers on the queue at each time period. The results are shown in Figure 14. Across all the scenarios, the run time of the proposed heuristic per iteration falls between 818 sec and 2126 sec, which are well within the simulation interval of 1 hour. All cases tend to show non-EV being less costly, which makes sense, while ChargerChasing EV case provides a good worst case bound among the scenarios.

Comparing operating strategies, the waiting time histogram in Figure 14(a) illustrates the inefficiency of the ChargerChasing heuristic since many customers never get served due to the immediate assignment of idle vehicles to charging stations. The current no-rebalancing setting also fails to guarantee a minimum service time for users. On the other hand, the non-myopic heuristic with $\eta = 0.85$ and $\beta = 1$ results in more evenly distributed waiting times. Figure 14(b) shows the distribution of queue length to be weighted more towards zero in the non-myopic EV case than the myopic EV case. Figure 14(c) shows the non-myopic EV case having a longer rebalancing time than myopic case due to more strategic allocations. However, this additional cost is more than offset by the benefit in user cost savings, as indicated in the Total Cost row in Table 7 where non-myopic EV (37.8% reduction from myopic EV). The performance of the online version of the EV non-myopic algorithm (under the current programmed software) is also quite satisfactory ($-29\%$ from the EV myopic case, compared to $-38\%$ with the near instantaneous version of the algorithm) and demonstrates the appropriateness of the heuristic rebalancing algorithm in an online setting even under the prototype coding in MATLAB used. The non-EV no rebalance scenario has the lowest total cost, even more so than the non-EV myopic scenario, which is consistent with earlier literature (e.g. Chow and Regan, 2011) that shows myopic strategies can be costly to implement.

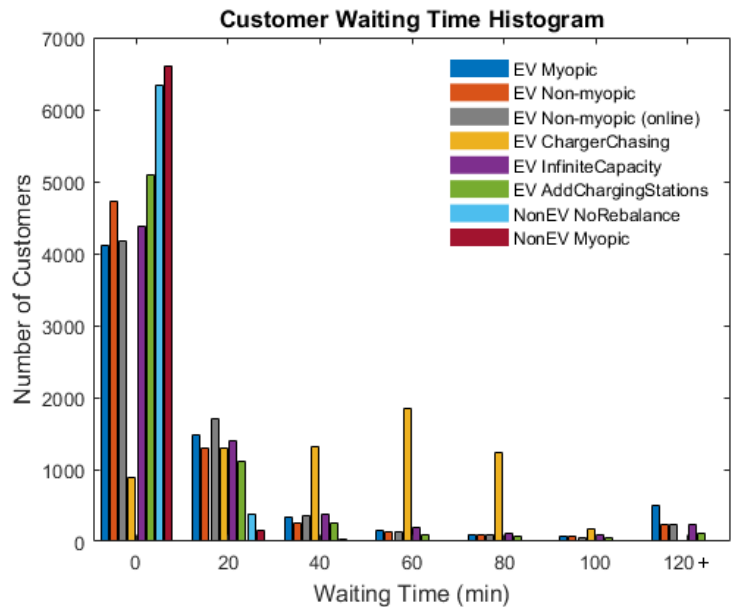

**Figure 14(a). Experienced waiting times**

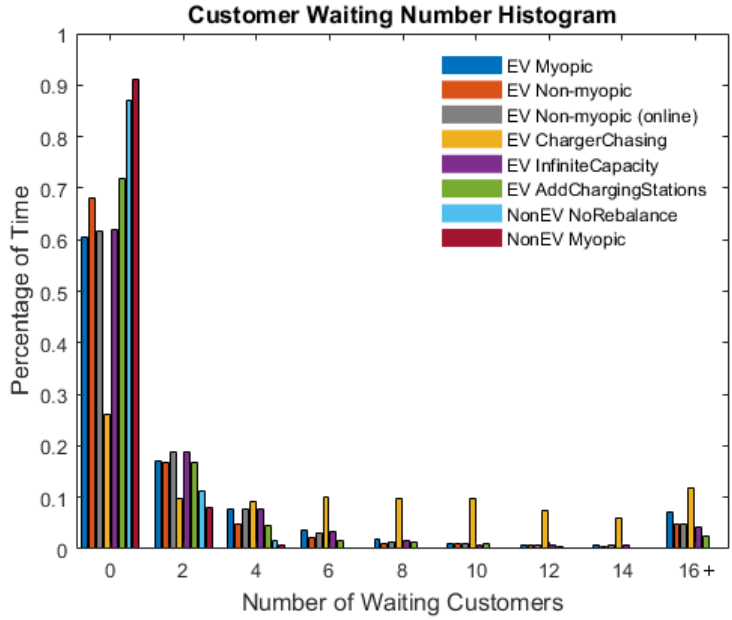
**Figure 14(b). Customer queue lengths**

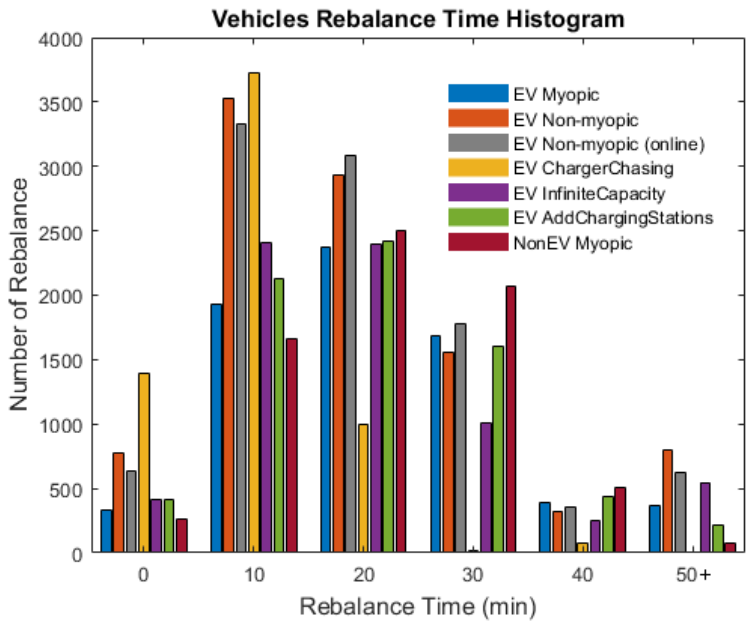
**Figure 14(c). Rebalance time per period**

**Figure 14.** Histograms of the performance metrics for the scenarios in the Brooklyn case study.

Other measures are summarized in Table 7. The total cost is given as the sum of realized waiting time and rebalance distance (which does not include recharging time) in the simulation, weighted by parameter $\theta = 0.02$ as defined in the MILP in Eq. (3) – (18). Computation time describes the average duration of the heuristic per rebalancing period. The average runtimes were obtained after "warm-starting" the simulation.

**Table 7.** Performance summary of various rebalancing strategies tested in the Brooklyn dataset

| Scenarios | EV Myopic | EV Non-Myopic | EV Non-Myopic (online) | Charger Chasing | Infinite Capacity | Add Charging Stations | Non-EV No Rebalance | Non-EV Myopic |
|---|---|---|---|---|---|---|---|---|
| Avg waiting time per customer (min) | 26.7 | 15.5 | 18.1 | 46.1 | 17.1 | 11.2 | 2.6 | 1.7 |
| Avg queue length per interval | 4.3 | 2.6 | 3.0 | 7.4 | 2.8 | 1.9 | 0.6 | 0.4 |
| Avg time per rebalance (min) | 26.7 | 70.0 | 39 | 10.6 | 22.8 | 23.4 | - | 22.2 |
| Rebalance number | 7073 | 9913 | 9807 | 6208 | 7026 | 7218 | - | 7086 |
| Computation time per interval (sec) | 1247 | 1074 | 914 | - | 817 | 1201 | - | 1169 |
| Total cost (Delay + θ×RebDis) | 190.4 | 118.4 (-38%) | 135.4 (-29%) | 316.7 (+66%) | 125.5 (-34%) | 85.2 (-55%) | 17.6 (-91%) | 20.6 (-89%) |

For the other scenarios, adding more capacity to the charging stations for this case study does not appear to matter much (and in fact resulted in marginally worse total cost in the simulation outcome in Table 7), which suggests the current capacities per charging station suffice. Spatially increasing the number of DC fast charging stations by 28% (from 18 to 23) proved much more significant in the system performance. These findings demonstrate important managerial insights to carshare companies deciding:

- Whether to adopt an EV fleet for a given existing charging infrastructure (the difference in costs are measured);
- Where to place charging infrastructure to best improve the operation of the system (different scenarios can be analyzed);
- How to operate a rebalancing policy that takes demand uncertainty into account (different algorithm parameters can be evaluated).

The heatmaps in Figure 15 provide an insight on the spatial distribution of available vehicles for four characteristic scenarios. From the idle vehicle distribution, we can see that the ChargerChasing strategy allocates vehicles in specific charging stations that are "conveniently" located in midpoint areas of the map. On the opposite spectrum, the scenario that includes the addition of charging stations leads to a sparser vehicle distribution.

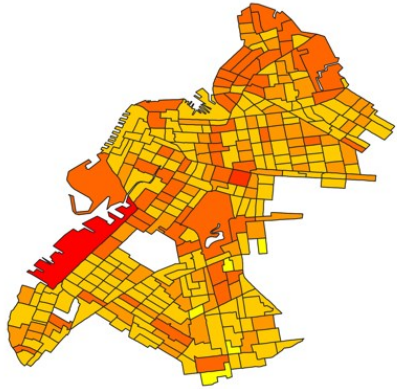
**Figure 15(a).** Myopic scenario

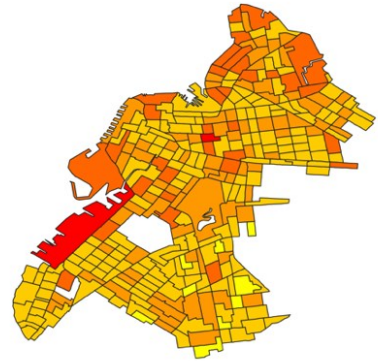
**Figure 15(b).** Non-Myopic

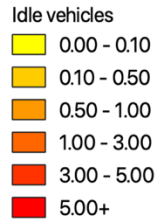

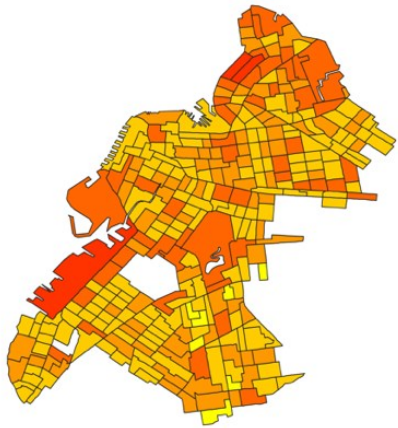
**Figure 15(c):** Additional charging stations

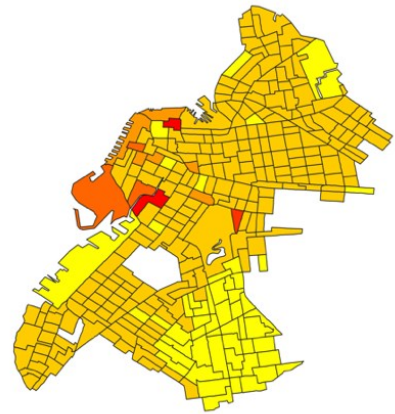
**Figure 15(d):** ChargerChasing

**Figure 15.** Average vehicles available in each TAZ zone heatmap

## 6. Conclusion

We propose a non-myopic idle vehicle rebalancing policy and greedy heuristic for an electric carsharing system that considers capacitated charging station constraints. To the best of our knowledge, this is the first facility relocation model formulation that considers queueing constraints applicable to EV charging. We formulate the cost function in each rebalancing time interval as a p-median problem embedded with a capacitated minimum cost flow network problem on a node-charge graph to jointly determine the relocation and routing decisions of idle vehicles under available charging capacity. The formulation on a two-dimensional node-charge graph (instead of a more computationally prohibitive node-time-battery graph) allows us to explicitly consider a customer's charging demand profile and optimize rebalancing operations of idle vehicles accordingly in an online system. An illustrative example on a small network shows the assigned vehicle flow and rebalanced positions of idle vehicles minimize total designed objective while optimally utilizing the available charging capacity in the system.

To address the computational complexities of real car-sharing networks, we propose a greedy heuristic algorithm that incorporates queuing constraints and solves the relocation problem in 15 – 89% of the computational time of commercial solvers for the MILP with only 7 – 35% optimality gaps in a single rebalancing decision time period. The online policy is tested on ten randomly generated test instances shows computational time reduction of 46% with an average gap of 58%. This allows the MDP heuristic to be operated in an online system.

From the large-scale Brooklyn simulation (262 vehicle fleet, 231 vehicle pickups per day, 303 TAZs, 5 charge levels), the proposed non-myopic algorithm can mitigate the cost of EV operations from a myopic algorithm relative to the non-EV no rebalance strategy by 38%. The online implementation of the same algorithm shows only a minor degradation in performance (-29% instead of -38%) establishing the heuristic's applicability in online settings.

We focused mainly on developing theoretical models for rebalancing electric vehicles as well as scalable heuristic algorithms for real-time operations. However, there is still more work to be done to evaluate and calibrate these theoretical models to real car-sharing operations. Potential stakeholders should carefully assess the benefits of using EV's against gasoline-fueled vehicles. The following list represents future research directions:

- Include dynamic demand (function of time, price and other factors)
- Include data-driven (machine learning) algorithms for updating $\lambda$
- More realistic/ commercial simulation environment
- Use data from larger operations
- Conduct detailed cost-benefit analysis on the tradeoffs of EV's and regular vehicles.

## Acknowledgements

This research was supported by the C2SMART University Transportation Center (USDOT #69A3551747124). One of the authors was supported by the Luxembourg National Research Fund (INTER/ MOBILITY/17/ 11588252). This work was also supported in part by the New York University Abu Dhabi (NYUAD) Center for Interacting Urban Networks (CITIES), funded by Tamkeen, through the New York University Abu Dhabi (NYUAD) Research Institute Award under Grant CG001, and in part by the Swiss Re Institute through the Quantum Cities$^{TM}$ Initiative. Data was provided by BMW ReachNow car-sharing operations in Brooklyn, New York, USA.## References

Barth, M. and Todd, M. (1999). Simulation model performance analysis of a multiple station shared vehicle system. *Transportation Research Part C* 7(4), 237–259.

Boyacı, B., Zografos, K. G. and Geroliminis, N. (2015). An optimization framework for the development of efficient one-way car-sharing systems. *European Journal of Operational Research* 240(3), 718–733.

Brandstätter, G., Kahr, M. and Leitner, M. (2017). Determining optimal locations for charging stations of electric car-sharing systems under stochastic demand. *Transportation Research Part B* 104, 17–35.

Bruglieri, M., Colorni, A. and Luè, A. (2014). The vehicle relocation problem for the one-way electric vehicle sharing: an application to the Milan case. *Procedia - Social and Behavioral Sciences* 111, 18–27.

Chargehub (2019). https://chargehub.com/en/charging-stations-map.html, last accessed Nov 9, 2019.

Chow, J. Y. J. (2019). BMW project C2SMART, https://zenodo.org/record/3333006.

Chow, J. Y. J., & Regan, A. C. (2011). Resource location and relocation models with rolling horizon forecasting for wildland fire planning. *INFOR: Information Systems and Operational Research*, 49(1), 31-43.31